\documentclass[runningheads]{llncs}
\usepackage{tabu}
\usepackage{graphicx}
\usepackage{hyperref}
\usepackage{todonotes}

\usepackage{amssymb}
\usepackage{amsmath}
\usepackage{wasysym}

\usepackage{subcaption}
\usepackage{xcolor}
\usepackage{soul}

\definecolor{clr}{rgb}{0.0, 0.0, 0.7}
\newcommand{\vorschlag}[1]{{{#1}}}

\begin{document}
\title{Scalable SAT Solving in the Cloud\thanks{This project has received funding from the European Research Council (ERC) under the European Union’s Horizon 2020 research and innovation programme (grant agreement No. 882500). 
		This work was performed on the supercomputer ForHLR funded by the Ministry of Science, Research and the Arts Baden-Württemberg and by the Federal Ministry of Education and Research.}}
\titlerunning{Scalable SAT Solving in the Cloud}
%
\author{Dominik Schreiber \and Peter Sanders}
\authorrunning{D. Schreiber and P. Sanders}
%
\institute{
	Karlsruhe Institute of Technology\\
	\email{\{dominik.schreiber,sanders\}@kit.edu}}
\maketitle              
\begin{abstract}
Previous efforts on making Satisfiability (SAT) solving fit for high performance computing (HPC) have lead to super-linear speedups on particular formulae, but for most inputs cannot make efficient use of a large number of processors. Moreover, long latencies (minutes to days) of job scheduling make large-scale SAT solving on demand impractical for most applications.
We address both issues with \textit{Mallob}, a framework for job scheduling in the context of SAT solving which exploits \textit{malleability}, i.e., the ability to add or remove processing power from a job during its computation. Mallob includes a massively parallel, distributed, and malleable SAT solving engine based on Hordesat with a more succinct and communication-efficient approach to clause sharing and numerous further improvements over its precursor. For example, Mallob on 640 cores outperforms an updated and improved configuration of Hordesat on 2560 cores.
Moreover, Mallob can also solve many formulae in parallel while dynamically adapting the assigned resources, and jobs arriving in the system are usually initiated within a fraction of a second. 

\end{abstract}
%
%
%

\section{Introduction}

Today's applications of SAT solving are manifold and include areas such as cryptography \cite{massacci2000logical}, formal software verification \cite{buning2019using}, and automated planning \cite{schreiber2021lilotane}.
Application-specific SAT encoders generate formulae which represent the problem at hand stated in propositional logic.
Oftentimes, multiple formulae which represent different aspects or horizons of the problem are generated \cite{buning2019using,schreiber2021lilotane}.
The difficulty of these individual formulae ranges from trivial to extremely difficult and is usually not known beforehand.
\vorschlag{Up to a certain degree, today's modern high performance computing (HPC) environments can facilitate the resolution of difficult problems.
In particular, we notice increased interest in on-demand SAT solving in cloud environments \cite{ngoko2017distributed,heisinger2020distributed} which is also reflected in the previous International SAT Competition \cite{balyo2020proceedings}.}
However, prior achievements of superlinear speedups for particular application instances \cite{balyo2015hordesat} must be set in relation with the total work which must be invested in every single formula to achieve such peak speedups.
Furthermore, in most HPC systems, long latencies of job scheduling (ranging from minutes to days) hinder the quick resolution of a stream of jobs even if most of the jobs are trivial.
To address these issues, we believe that a SAT solver launched on a formula of unknown difficulty should be allotted a flexible amount of computational resources based on the overall system load and further task-dependent parameters.
In the context of scheduling and load balancing, this feature is called \emph{malleability}: The ability of an algorithm to deal with a varying number of processing elements throughout its computation.
Malleable algorithms open up opportunities for highly dynamic load balancing techniques: The number of associated processing elements for each job can be adjusted continuously to warrant optimal and fair usage of available system resources.




In this work, \vorschlag{we outline a randomized and decentralized framework \textit{Mallob} for malleable scheduling and load balancing of SAT jobs; we present a new large-scale SAT solving system that is both fully distributed and malleable; and we connect these components to obtain a malleable scheduling platform for the scalable resolution of SAT jobs.}
For this means, we revisit the popular massively parallel SAT solver Hordesat \cite{balyo2015hordesat} and carefully re-engineer most of its components. 
\vorschlag{Most notably, we propose a succinct and communication-efficient clause exchange mechanism, update and adapt Hordesat's solver backend to support malleability, and integrate practical performance improvements.}
Evaluations show that our solver on 32 nodes (640~cores) outperforms an improved configuration of Hordesat on 128 nodes (2560~cores) when both employ the same portfolio of solvers, and our solver consistently shows improved speedups.
\vorschlag{
Mallob's job scheduling and load balancing on top of this solver imposes minimal overhead and can combine parallel job processing with a flexible degree of conventional parallel SAT solving to make best use of the available resources.
In most cases it only takes a fraction of a second until an arriving job is initiated.
}

After describing important preliminaries and related work in Section~\ref{sec:related-work}, we present the malleable environment which hosts our solver engine in Chapter~\ref{sec:malleable-environment}.
Thereupon, in Chapter~\ref{sec:solver} we present the solver engine itself.
We present the evaluation of our solver in Chapter~\ref{sec:evaluation} and conclude our work in Chapter~\ref{sec:conclusion}.

\section{Related Work}
\label{sec:related-work}


For the sequential resolution of SAT problems, the most commonly used algorithm is \textit{CDCL} \cite{marques2009conflict} which is essentially a highly engineered heuristic depth-first search over the space of partial variable assignments.
CDCL features numerous advanced techniques such as non-chronological backtracking and restart mechanisms.
Most notably for our work, when a logical conflict is encountered during search, a clause which represents this conflict can be learned and added to the solver's clause database.
The additional knowledge gained from this learning mechanism can help to speed up the subsequent search.
Another branch of notable sequential SAT solving approaches is the family of \textit{local search solvers} which perform stochastic local search over the space of variable assignments \cite{hoos2000local}.

SAT solving in parallel is mostly based on using existing sequential SAT solvers as building blocks.
One strategy which is often called the \textit{portfolio approach} is to execute several solvers
on the same formula, e.g., \cite{hamadi2010manysat,audemard2014dolius}.
Diversification strategies for an effective portfolio range from
supplying different random seeds to the same solver over reconfiguring the solver's parameters
to employing wholly different SAT solvers.
As an alternative to portfolio approaches, \textit{search space partitioning approaches} subdivide the original formula into several sub-formulae and solve these in parallel, e.g., \cite{schubert2010pamiraxt,audemard2016adaptive}.
An extreme case of this strategy is applied in parallel \textit{Cube\&Conquer} approaches: C\&C generates a large number (potentially millions) of subproblems using so-called lookahead solvers and then distributes the problems among all workers \cite{heule2011cube,heisinger2020distributed}.
Regardless of the means of parallelization, an important feature of parallel solvers is to exchange learnt clauses among all workers and, notably, to find a good tradeoff between the sharing of useful information and the avoidance of unnecessary overhead \cite{ehlers2014communication}. 

The International SAT Competition 2020 established a distinction between modestly parallel SAT solving and HPC SAT solving by featuring both a parallel track and a cloud track.
The parallel track was evaluated on a single node with 32 cores and a time limit of 5000s per instance while the cloud track was evaluated on 100 nodes with 8 cores each and a considerably lower time limit of 1000s per instance.
These different modes of operation have far-reaching consequences on the design of solvers:
For modest levels of (shared-memory) parallelism, high memory consumption and high concurrency can become a considerable issue \cite{iser2019memory}.
In a large-scale cloud environment, shared-memory concurrency can be less of an issue while good diversification and communication efficiency becomes critical.

Hordesat \cite{balyo2015hordesat} is an example for a SAT solver specifically designed for massive parallelism.
It features a modular solver interface which allows to plug in and dynamically diversify different SAT solvers without changing their internal workings.
Clause exchange is performed in discrete \textit{rounds} via all-to-all collective operations.
The Hordesat paradigm found adoption in a generic interface for parallel SAT solving \cite{le2017painless}.
Hordesat was also used as the baseline example for the setup of the cloud track of the SAT Competition 2020 \cite{balyo2020competition}.
Our solver system is based on Hordesat and scored the first place in this competitive event. 

Previously, a distributed system for SAT solving in the cloud was presented in \cite{ngoko2017distributed,ngoko2019solving}.
\vorschlag{It features a centralized malleable scheduler which precomputes a schedule based on run time predictions and which employs sequential solvers without any communication among them: The authors noted that ``\textit{such solutions [for exchange of knowledge] are not necessarily suitable for distributed clouds in which the communication time could be important}'' \cite{ngoko2017distributed}.
In contrast, we demonstrate that clause exchange is highly effective and introduce a decentralized scheduling approach with dynamic load balancing that does not rely on any run time predictions.}
Another work related to ours is the distributed Cube\&Conquer solver Paracooba \cite{heisinger2020distributed}.
It also supports parallel processing of multiple jobs and performs malleable load balancing.
While Paracooba is designed for Cube\&Conquer, \vorschlag{we propose a malleable portfolio approach. 
Regarding SAT solving performance, our system clearly outperformed Paracooba in the past competition \cite{balyo2020competition}.}

\section{Malleable Environment}
\label{sec:malleable-environment}

We now outline the platform \textit{Mallob} for the scheduling and load balancing of malleable jobs.
{Mallob} is an acronym of \textbf{Mal}leable \textbf{Lo}ad \textbf{B}alancer as well as \textbf{M}ulti-tasking \textbf{A}gi\textbf{l}e \textbf{Lo}gic \textbf{B}lackbox.
As a comprehensive presentation of Mallob in its entirety is too broad in scope for this publication, we present the design decisions and the features of Mallob that are necessary to understand our SAT solving system and will describe the internal workings and theoretical properties of our scheduling and load balancing in a future publication.

\begin{figure}
	\centering
	\includegraphics[width=0.8\textwidth]{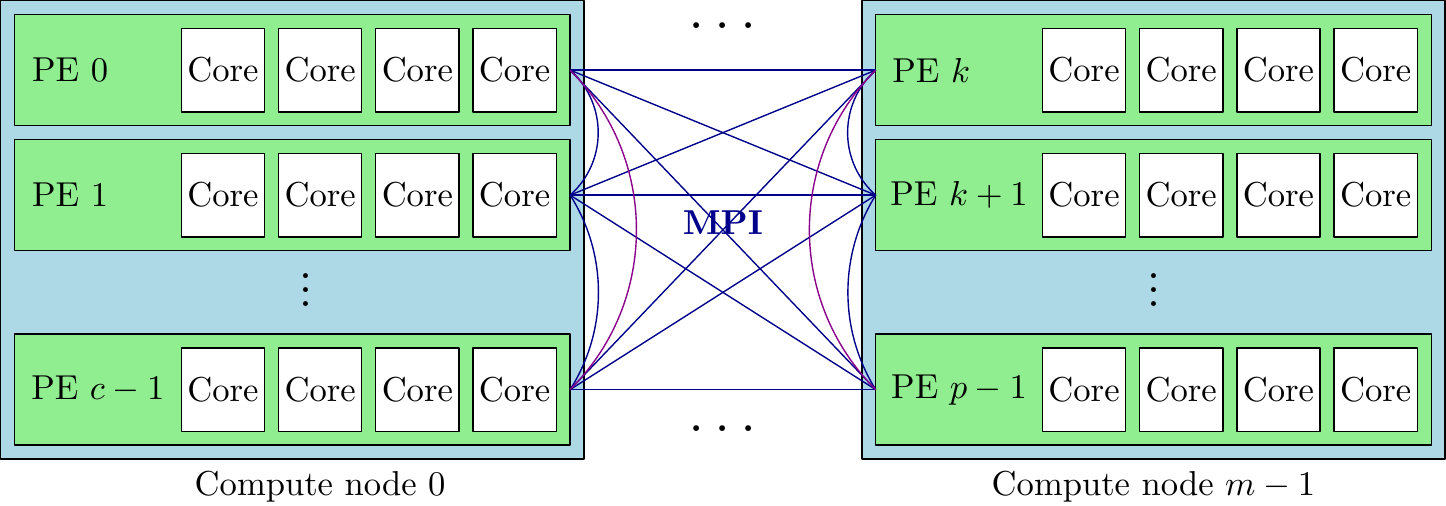}
	\caption{System architecture used by Mallob}
\end{figure}

We consider a homogeneous\footnote{\vorschlag{While we intend to generalize our system to heterogeneous environments in the future, this undertaking is out of scope for this publication.}} 
distributed computing environment with $m$ \emph{compute nodes}.
For the sake of generality, we do not assume any kind of shared (RAM or disk) memory between the compute nodes.
As such, the only way for the nodes to exchange information is to send messages over some broadband interface.
This is enabled by the Message Passing Interface (MPI) \cite{gropp1999using}.

Each compute node contains several \emph{cores}.
  We partition the cores on a node
  into $c$ \emph{groups} of $t$ cores each running one \emph{thread}.%
  \footnote{The cores may be distributed over several CPU chips (or
    sockets). Moreover, each core may be able to run several hardware
    threads
    . Our system can handle both additional levels of hierarchy by appropriately defining $c$ and $t$ but we abstract from this in the remainder of this paper.}
  Each group is implemented as a \emph{process} and is also called \emph{PE} (for \emph{processing element}) in the following.
Overall, our system contains a total of $p := c \cdot m$ PEs and $c\cdot m\cdot t$ threads.

A number of \textit{jobs} $1, \ldots, n$ arrive in the system at arbitrary times. 
A job is a particular problem statement, in our case given by a propositional logic formula in \textit{Conjunctive Normal Form} (CNF).
Every job $j$ has a constant \textit{priority} $\pi_j \in (0, 1)$ and a \emph{demand of resources} $d_j \in \mathbb{N}$ which may vary over time.
In the most simple setting, $d_j = p$ at all times.
More generally, a job can express with $d_j$ how many PEs it is able to employ in its current stage of computation.
We expect the number of jobs to be smaller than the number of workers, $n < p$, which allows us to restrict each PE to compute on at most one job at a time.


If a job $j$ enters the system from an external interface, then a request message $r_0(j)$ will perform a random walk through a sparse regular graph over all PEs until an idle PE $p_0(j)$, named the \textit{root} of $j$, adopts the job.
This root remains unchanged throughout the job's lifetime and represents $j$ in collective load balancing computations.
\vorschlag{Such a balancing computation is triggered at most once within a certain period $e$ (e.g., $e=0.1s$).
Possible events which trigger a balancing are (a) the arrival of a new job, (b) the completion of a job, and (c) the change of a job's demand.
These events are then broadcast globally with a single lightweight collective operation.}
The result of each balancing is a map $j \mapsto v_j$ which assigns to each job $j$ a certain integer, the \emph{volume} $v_j \geq 0$. 
$v_j$ is proportional to $d_jp_j / \sum_{j'} d_{j'}p_{j'}$ and determines the number of PEs which are supposed to compute on $j$ until the next update of $v_j$.

\label{sec:job-trees}

\begin{figure}[t!]
	\centering
	\includegraphics[width=0.9\textwidth]{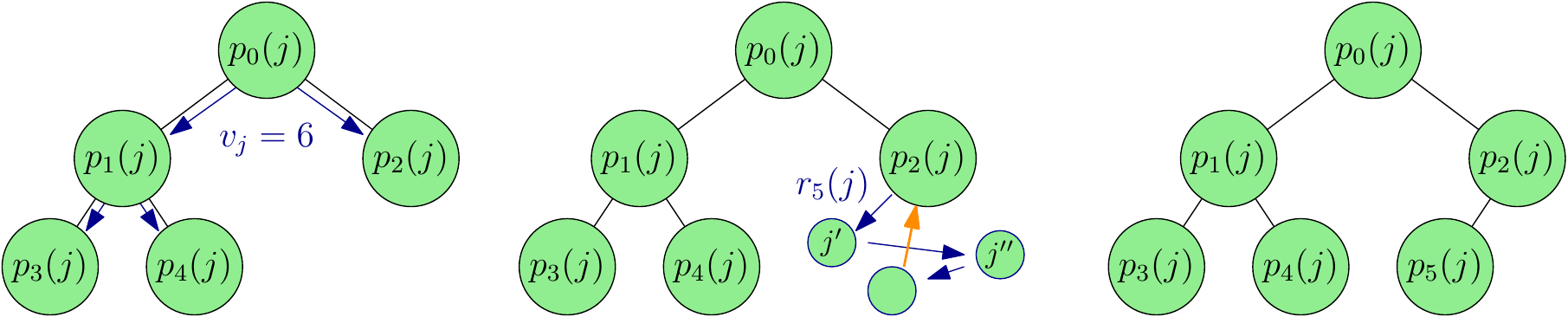}
	\caption{Illustration of $T_j$ growing from volume 5 to 6. Each circle is a PE.}
\end{figure}




The \textit{job tree} $T_j$ of job $j$ is a binary tree of PEs that is rooted at $p_0(j)$.
Its purpose is to enforce the volume assigned to $j$ and to enable efficient job-internal communication.
Each node $p_x(j)$ in $T_j$ has a unique index $x \geq 0$.
Node $p_x(j)$ may have child nodes $p_{2x+1}(j)$ (left child) and $p_{2x+2}(j)$ (right child).
$T_j$ is supposed to consist of exactly $v_j$ nodes $p_0(j), \ldots, p_{v_j-1}(j)$ and adjusts accordingly whenever $v_j$ updates:
Beginning from $p_0(j)$ which computes an update, an according message is recursively broadcast over $T_j$.
If this volume update arrives at a node $p_x(j)$ for which $x \geq v_j$, then the node will leave $T_j$ and suspend its solvers.
Conversely, if $p_x(j)$ does not have a left (right) child node and if $2x+1 < v_j$ ($2x+2 < v_j$), it will send out an according request for another idle PE to join $T_j$.
These messages are first routed over any former children of $p_x(j)$ before they begin a random walk.
As such, our node allocation strategy prioritizes PEs which may still host suspended solvers for $j$.
\vorschlag{In order to make careful use of main memory, we allow each PE to host a small constant number of job nodes and to discard the oldest job nodes if this limit is exceeded.}

\section{The Mallob SAT Engine}
\label{sec:solver}

We now present our massively parallel, distributed, and malleable SAT solving system named the \textit{Mallob SAT Engine}.
\vorschlag{We focus on the following three points: (1) A careful clause exchange which only transmits globally useful data and fits our malleable computation model; (2) a rework of Hordesat's solver backend to support malleability in a natural manner and to render it competitive; and
(3) practical optimizations and performance improvements over Hordesat.}

\subsection{\vorschlag{Succinct Clause Exchange}}


Hordesat uses synchronous communication in \textit{rounds} to periodically perform an all-to-all clause exchange.
The used collective operation is called an \textit{all-gather}: Each PE $i$ contributes a buffer $b_i$ of fixed size $\beta$.
\vorschlag{Each buffer contains a serialization of a set of learned clauses which were \textit{exported} by the PE's solver threads.
The first integer $n_1$ in $b_i$ denotes the number of clauses of size 1, followed by $n_1$ integers representing these clauses.}
The next integer $n_2$ denotes the number of clauses of size 2, followed by $2 \cdot n_2$ integers representing these clauses; and so on.
The concatenation of all buffers, $B := \langle b_1, \ldots, b_p \rangle$, is then broadcast to each PE.
This all-gather operation is included by default in all MPI implementations. 
\vorschlag{Then each solver \textit{imports} shared clauses into its individual database of clauses.}

We noticed that the above clause exchange mechanism has various shortcomings. 
First, whenever a node does not completely fill its local clause buffer, the broadcast data contains ``holes'' which carry no information but are nevertheless sent around.
Secondly, the broadcast clause buffer may contain a significant portion of redundant clauses:
In particular in the beginning of SAT solving when a formula is simplified and basic propagations are done, we noticed that this can lead to $p$ very similar sets of clauses within $B$.
This effect is especially pronounced for unit clauses which are never filtered out (see below).
Thirdly, the length of the message that is broadcast to all nodes grows proportionally to the number of involved PEs.
For sufficiently large Hordesat configurations, this can constitute a bottleneck both in communication volume and in the local work necessary to digest and process every clause received.

\begin{figure}[b!]
	\centering
	\includegraphics[width=0.7\textwidth]{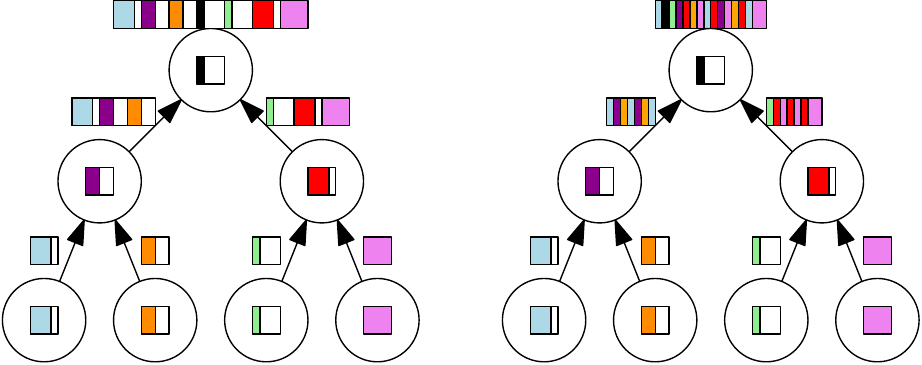}
	\caption{Exemplary flow of information in the first half of Hordesat's all-gather operation (left) and in our aggregation within a job tree (right).
	Each circle is a PE; a buffer within a circle represents the PE's locally collected clauses.} \label{fig:clauseexchange}
\end{figure}

In our system, the set of active PEs of a job $j$ is represented by a job tree $T_j$ as described in Section~\ref{sec:job-trees}.
We can conveniently use $T_j$ as the communication structure for our clause exchange 
and therefore ensure that the set of PEs which participate in the collective operation is equivalent to the set of PEs that are currently affiliated with the job.
\vorschlag{As soon as a fixed amount of time $s$ has passed since the last broadcast of shared clauses (e.g., $s=1$~second), each leaf node in $T_j$ send its exported clauses to its parent.}
When an inner node has received as many buffers as it has children, \vorschlag{it exports its own clauses} and then performs a two- or three-way merge of the present buffers up to a certain limit:
All input buffers are read simultaneously from left to right and aggregated into a single new buffer, similar to the merge of sorted sequences in textbook Mergesort.
A hash set of seen clauses with commutative hashing \cite{balyo2015hordesat} allows us to immediately recognize and filter duplicate clauses during the merge. 
The size of the new buffer is limited and any remaining unread information in the input buffers is discarded.
\vorschlag{As each clause buffer is sorted in increasing order by clause length after each merge, we aggregate some of the globally most valuable information while we strictly limit the overall communication volume.
Furthermore, we improve the density of useful information in the transferred data because each sent buffer is of compact shape and contains no redundant clauses.}

\vorschlag{We chose the limit on the buffer size for $p_x(j)$ as follows:
Bundled with each buffer payload we communicate the number $u$ of buffers aggregated so far.}
For each aggregation step, i.e., for each further level of $T_j$ that is reached, we want to discount the maximum buffer size by a factor of $\alpha$.
As a consequence, we compute the buffer size limit $b(u) := \lceil u \cdot \alpha^{\log_2(u)} \cdot \beta \rceil$.
This number of shared clause literals can be steered by a user parameter $\alpha \in [\frac{1}{2}, 1]$, the \emph{discount factor} at each buffer aggregation.
For $\alpha = \frac{1}{2}$, we can see that the clause buffer size converges to $\beta$; for $\alpha > \frac{1}{2}$, the clause buffer size diverges.
For $\alpha = 1$, our approach emulates Hordesat's shared clause buffer which grows proportionally without any discount.

\label{sec:clause-filtering}

Hordesat employs \textit{clause filtering} to detect and discard redundant clauses which have already been imported or exported before.
\vorschlag{This technique is realized with 
an approximate membership query (AMQ) data structure on the basis of commutative clause hashing \cite{balyo2015hordesat}.}
Each PE of Hordesat employs one \textit{node filter} and $t$ \textit{solver filters} (one for each solver thread).
At clause export, each clause is registered in its solver filter and then tested against the node filter.
At clause import, each clause is tested against the node filter and then against each solver filter.
\vorschlag{Unit clauses, however, are always admitted due to their relative importance.
This is problematic because particular unit clauses can be sent around many times and can waste considerable amounts of space in the buffers.}

We omitted node filters because their main use is to check for duplicate clauses, what is already done during the aggregation of buffers in our case.
\vorschlag{We complemented the solver filters with an additional filtering of unit clauses, using an exact set instead of an AMQ data structure.}
This way we do not get any false positives for unit clauses and make sure that each such clause is being shared.
Last but not least, we implemented a mechanism similar to restarts into the clause filters:
\vorschlag{Every $X$ seconds, 
half of all clauses (chosen randomly) in each clause filter are forgotten and therefore can be shared again. 
This mechanism also enables solvers to learn crucial clauses at a later point even if they (re-)join $T_j$ after these clauses have already been shared for the first time.}

\subsection{\vorschlag{Malleable Solver Backend}}

In the following we present the most relevant changes we made to Hordesat's solver backend to naturally support malleability.

\subsubsection{Diversification.}

\vorschlag{As in Hordesat, our approach relies on three different sources of diversification: Employing different solver configurations, handing different random seeds to the solvers, and supplying each solver with different default polarities (\textit{phases}) of variables.
We diversify a particular solver $S$ with a \textit{diversification index} $x_S \geq 0$ and a \textit{diversification seed} $\sigma_S$.
We use $x_S$ to determine a particular solver configuration, and we use $\sigma_S$ as a random seed and to select random variable phases.}
The $i$-th solver $S$ ($0 \leq i < t$) employed by $p_k(j)$ is assigned $x_S := kt + i$.
We obtain $\sigma_S$ by combining $x_S$ with the solver's thread ID (given by the operating system). As such, each instantiated solver has a slightly different diversification even if a job node is rescheduled and a solver $S'$ is instantiated for which some solver $S$ with $x_S = x_S'$ already existed before.

We introduce a number of updates to the portfolio of solvers compared to Hordesat.
We noticed that the clause sharing mechanism implemented in the Minisat interface of Hordesat is disruptive to the solving procedure: Whenever a certain threshold of collected clauses is reached, the solver is interrupted to add all clauses and then restarted.
This periodic interruption might have caused the bad performance of Hordesat with Minisat in the original evaluations.
For this work we focus on Lingeling as an efficient and reliable SAT solver with great diversification options.
We updated Lingeling from its 2014 version \cite{biere2014yet} to its 2018 version \cite{biere2017cadical} with the side effect of rendering all core modules of our system Free Software.
Similarly, instead of the 16 diversification options from the former Plingeling \cite{biere2014yet}, we use 13 CDCL diversification options from the newer Plingeling \cite{biere2017cadical}.
Every fourteenth solver thread uses local search solver YalSAT (included in the Lingeling interface), alternatingly with and without preprocessing.

\subsubsection{Preemption of Solvers.}

\vorschlag{To support malleability, it is essential that a PE's management thread can suspend, resume, and terminate each job node at will.
We noticed that we cannot rely on each solver thread periodically calling an according callback function because the solver can get stuck in expensive preprocessing and inprocessing \cite{biere2016splatz}.
To still enable seamless preemption, 
we enabled our solver engine to be launched in a separate process.}
This does involve some overhead as a new process is forked, a shared memory segment for efficient Inter-Process Communication (IPC) is set up, and an additional management thread in the SAT solving process is employed.
However, suspension and termination of a process is supported on the OS level in a safe and elegant manner through signals.
As solver threads may be unresponsive when the SAT engine's process catches a termination signal, they are interrupted and cleaned up forcefully.
Nevertheless, this leaves the main process and the system in a valid state.

\subsection{\vorschlag{Practical Improvements}}

\vorschlag{We now present some further practical improvements of Mallob over Hordesat.}

\subsubsection{Lock-free Clause Import.}

For each solver $S$ within a PE, the main thread of Hordesat copies all admitted clauses from clause sharing into a buffer $b_S$ and increases its size as necessary.
The solver thread of $S$ then imports the clauses in $b_S$ one by one.
As this implies a race condition, $b_S$ is guarded by a mutex which is locked by the solver thread before reading clauses and by the main thread before writing clauses.
If the solver thread cannot acquire this lock, it gives up on importing a clause.
We replaced $b_S$ with a lock-free ring buffer\footnote{https://github.com/rmind/ringbuf} $r_S$ 
and hence achieve a lock-free import of clauses while also making more careful use of the available memory: The size of $r_S$ is fixed and clauses are eventually discarded if a solver consumes no clauses for some time.
We set $|r_S|$ to a low multiple of the maximum number $L$ of literals which may be shared in a single round.

\subsubsection{Memory Usage.}

The memory consumption of parallel SAT solvers is a known issue \cite{iser2019memory}:
As each solver commonly maintains its own clause database, the memory requirements increase proportionally with the number of spawned solvers.
\vorschlag{As such, large formulae can cause Out-Of-Memory errors.}
To counteract this issue, we introduce a simple but effective step of precaution:
For a given threshold $\hat{s}$, if a given serialized formula description has size $s > \hat{s}$, then only $t' = \max\{1, \lfloor t \cdot \hat{s} / s \rfloor\}$ threads will be spawned for each PE.
The choice of $\hat{s}$ depends on the amount of available main memory per PE.
For the system we used, 3.2~GB per solver thread are available:
Based on monitoring the memory usage for different large formulae in our system, we use $\hat{s} := 10^8$.
\vorschlag{As $t'$ only depends on $s$, the $t'$ threads can be started immediately without any further inspection of the formula.}

\section{Evaluation}
\label{sec:evaluation}

We now turn to the evaluation of our work.
After explaining the experimental setup, we first evaluate the capabilities of our SAT solver engine, from now on denoted \textit{Mallob-mono} (as in the SAT Competition), in a standalone setting.
Then we evaluate Mallob in its entirety, i.e., with malleable job scheduling.

We implemented Mallob in C++17 and make use of OpenMPI \cite{graham2006open}.
Our software is available at \texttt{github.com/domschrei/mallob} and all experimental data is available at \texttt{github.com/domschrei/mallob-experimental-data}.

\vorschlag{We mostly tested the parallel solvers on a comparably low time limit of 300 seconds because we are interested in solving SAT instances as rapidly as possible in a distributed cloud environment.
Furthermore, we save resources with this lower time limit and can in turn test more different configurations.}

We evaluate Hordesat both with its original portfolio and with the updated portfolio that Mallob uses.
For both Hordesat configurations, we fixed a performance bug to make Hordesat more competitive:
\vorschlag{Lingeling repeatedly queries the elapsed time since its initialization.
In the original code, no callback providing this elapsed time was given to Lingeling which caused each solver thread to fall back to expensive system calls instead.
Depending on the configuration this lead to each solver spending more than 10\% of its time in kernel mode.}

We performed all experiments on the ForHLR phase II which is a general purpose high performance computing (HPC) cluster.
The ForHLR features 1152 compute nodes each with two 10-core Intel Xeon processors E5-2660 v3 clocked at 2.6~GHz.
Each node has 64~GB of main memory (RAM).
All nodes are connected by an InfiniBand 4X EDR interconnect.
We performed our experiments on up to 128 nodes with a total of 2560 physical cores.
Consistent with the default configuration of Hordesat, we bind each MPI process to four physical cores (or eight virtual cores).
Consequently, we execute $20/4=5$ MPI processes on each node which results in up to $128 \cdot 5 = 640$ PEs \vorschlag{with up to four solvers each}.

\subsection{Selection of Benchmarks}

For the unbiased evaluation of SAT solvers, a benchmark consisting of many diverse formulae from different origins is necessary.
However, the computational resources which are required to evaluate a software increase proportionally with the number of involved compute nodes.
In our massively parallel setting, we are committed to perform responsible and resource-efficient evaluations while still aiming for statistical relevance and robustness of results.

We analyzed the benchmarks of the SAT Competition 2020 with GBD \cite{iser2019problem} and partitioned them into 80 separate \textit{families} (including families from past competitions).
We sorted the instances of each family according to the number of contained clauses and then randomly picked one SAT instance from the second (larger) half of each family's sorted instance list.
As such, we obtained 80 instances (35 satisfiable, 35 unsatisfiable, 10 ``unknown'') which we believe to be a reasonably diverse and difficult set of benchmarks for our means.
\vorschlag{However, we are aware that the reduction of a test set generally increases the risk of overfitting.
As such, we only accept a more complicated configuration when we believe the difference in runtimes to be significant and reject it otherwise.
That being said, we did perform one comparison of the best known configurations of Hordesat and Mallob-mono on the entire set of 400 benchmarks (see Appendix).}

\subsection{SAT Solving Performance}

We now discuss the results of our evaluations.
We provide penalized average runtime (PAR-2) scores which penalize timeouts with twice the time limit and which were also employed in the SAT Competition 2020 \cite{balyo2020competition}.

First, let us take a look at the overall performance we measured for Hordesat and for Mallob(-mono) on 128 compute nodes.
We included Hordesat both with its original solvers (``old'') and with Mallob's updated portfolio (``new'').
We included Mallob in a basic configuration that is as close as possible to Hordesat, i.e., with discount factor $\alpha=1$.
Furthermore, Hordesat imposes an upper bound on the LBD or ``glue'' value \cite{audemard2009predicting} of clauses that are exported: Initially, a clause must be unit or have an optimal LBD score of 2 to be shared, and whenever a PE fills its clause buffer by less than 80\% the limit is incremented.
We also adopted this mechanism in this configuration.
In addition, we tested four more different values of $\alpha$ to explore the effectiveness of sublinear clause buffer growths.

\begin{figure}[t!]
	\centering
	\begin{subfigure}{0.52\textwidth}
		\includegraphics[width=\textwidth]{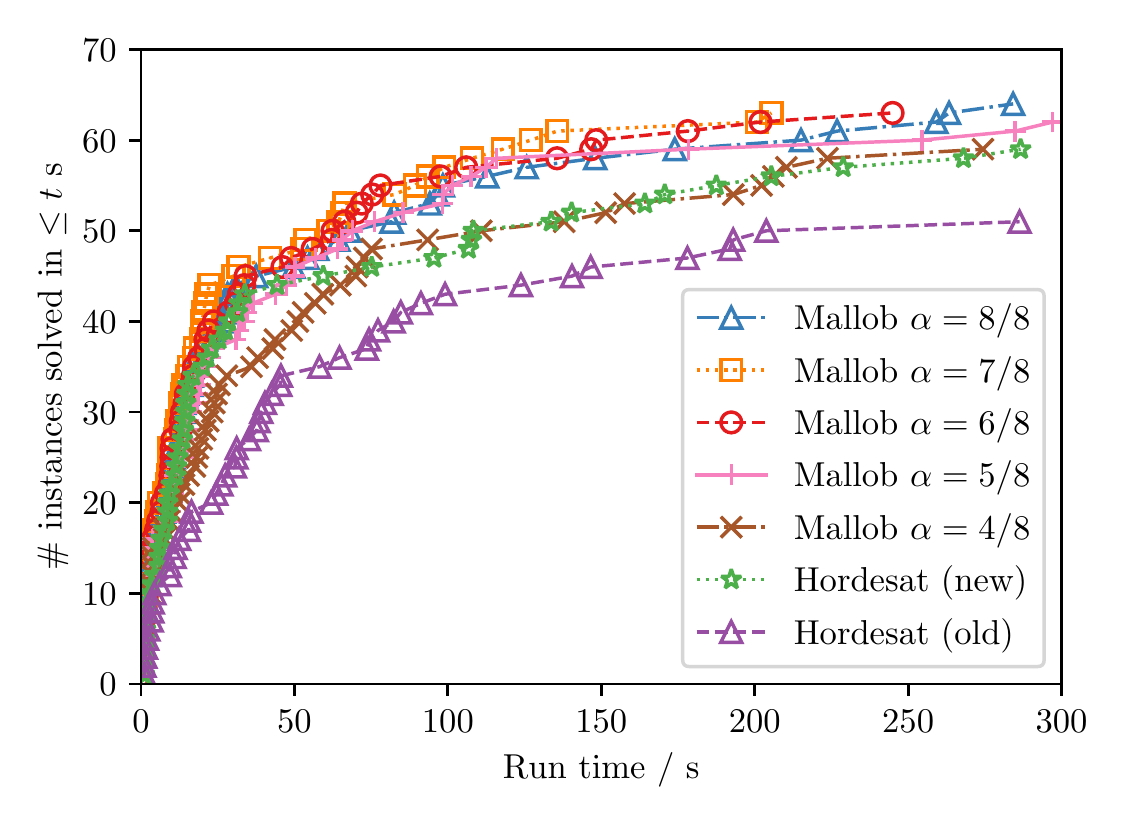}
	\end{subfigure}
	\begin{subfigure}{0.46\textwidth}
		\setlength\tabcolsep{1mm}
		\scriptsize \begin{tabular}{lr|rrrr}
			Configuration & \#I. & \# & (+, & -) & PAR-2 \\\hline
			Hordesat (old) & 80 & 51 & 23 & 28 & 252.7 \\
			Hordesat & 80 & 59 & 28 & 31 & 193.7 \\
			Mallob $\alpha=4/8$ & 80 & 59 & 29 & 30 & 196.2 \\
			Mallob $\alpha=5/8$ & 80 & 62 & 30 & 32 & 169.6 \\
			Mallob $\alpha=6/8$ & 80 & 63 & 30 & 33 & 157.4 \\
			Mallob $\alpha=7/8$ & 80 & 63 & 31 & 32 & \textbf{154.0} \\
			Mallob $\alpha=8/8$ & 80 & 64 & 31 & 33 & 158.2 \\
		\end{tabular}
	\end{subfigure}\vspace*{-3mm}
	\caption{\vorschlag{Performance of Hordesat (with original and updated solvers) and of ``na\"ive'' Mallob (with different values of $\alpha$) on 128 compute nodes. Table columns: Number of instances, solved instances (SAT, UNSAT), PAR-2 scores.}}
	\label{fig:horde-mallob-overview}
\end{figure}

As Fig.~\ref{fig:horde-mallob-overview} shows, the update of solver backends brings a clear improvement for Hordesat.
Furthermore, the most na\"ive and untuned configuration of Mallob with $\alpha = 1$ outperforms Hordesat even if both systems make use of the exact same backend solvers.
If $\alpha=0.5$, only a very small clause buffer of less than 1500 integers is shared in each round which proves to be highly detrimental to Mallob's performance and underlines the importance of clause sharing.
The best overall performance is achieved with $\alpha = 7/8$ whereas $\alpha = 6/8$ is a close second.

In a next step, we discuss different constraints on shared clauses we tested.
\vorschlag{The exact results are given in the Appendix (Table~\ref{tab:results}).
Experiments on 128 nodes revealed that the LBD limit mechanism of Hordesat which we used before is not beneficial to Mallob's performance; rather, best performance is achieved without such a limit.}
We also tested a maximum clause length limit of 5 and 10 and found the results to be mostly inconclusive.
As such, we arrive at a very simple configuration of Mallob without any additional limits on clause lengths or LBD scores.
For the clause filter half life $X$ we tested values of 10, 30, 90, and $\infty$ (i.e., no clauses are forgotten).
With $X=90$ three more instances were solved than with $X=\infty$, but the configuration with $X=\infty$ solved most instances slightly faster.
We continue the evaluation with $X=\infty$ \vorschlag{but will consider $X=90$ for Mallob's scheduling mode where the preemption of PEs and a longer wallclock time limit per instance merit a re-sharing of clauses.
}

\begin{figure}[t!]
	\includegraphics[height=0.43\textwidth]{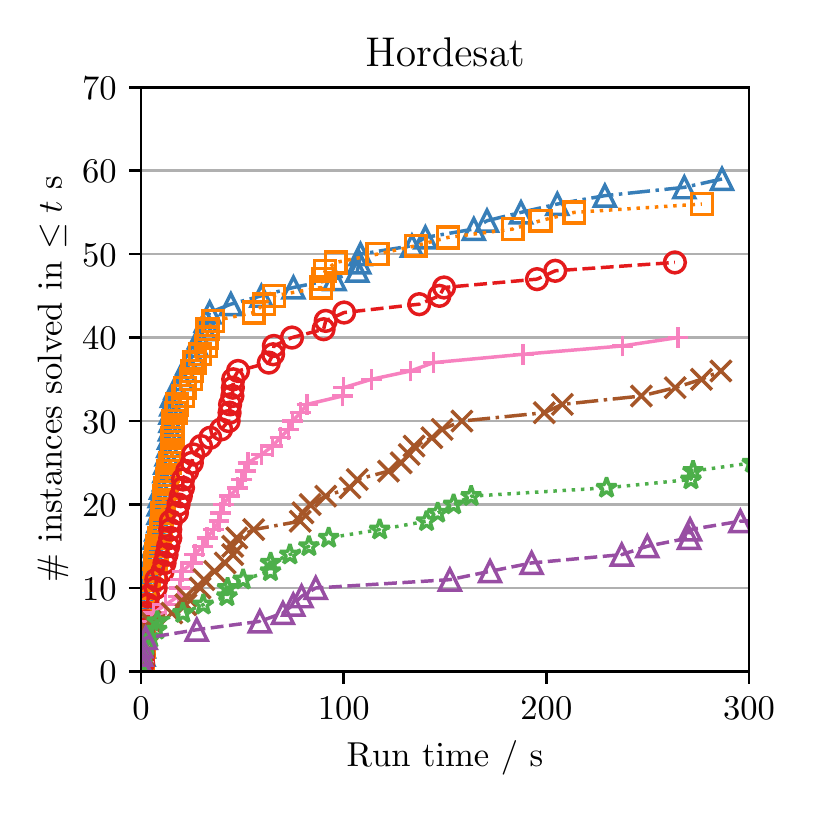}\hspace{-0.3cm}
	\includegraphics[height=0.43\textwidth]{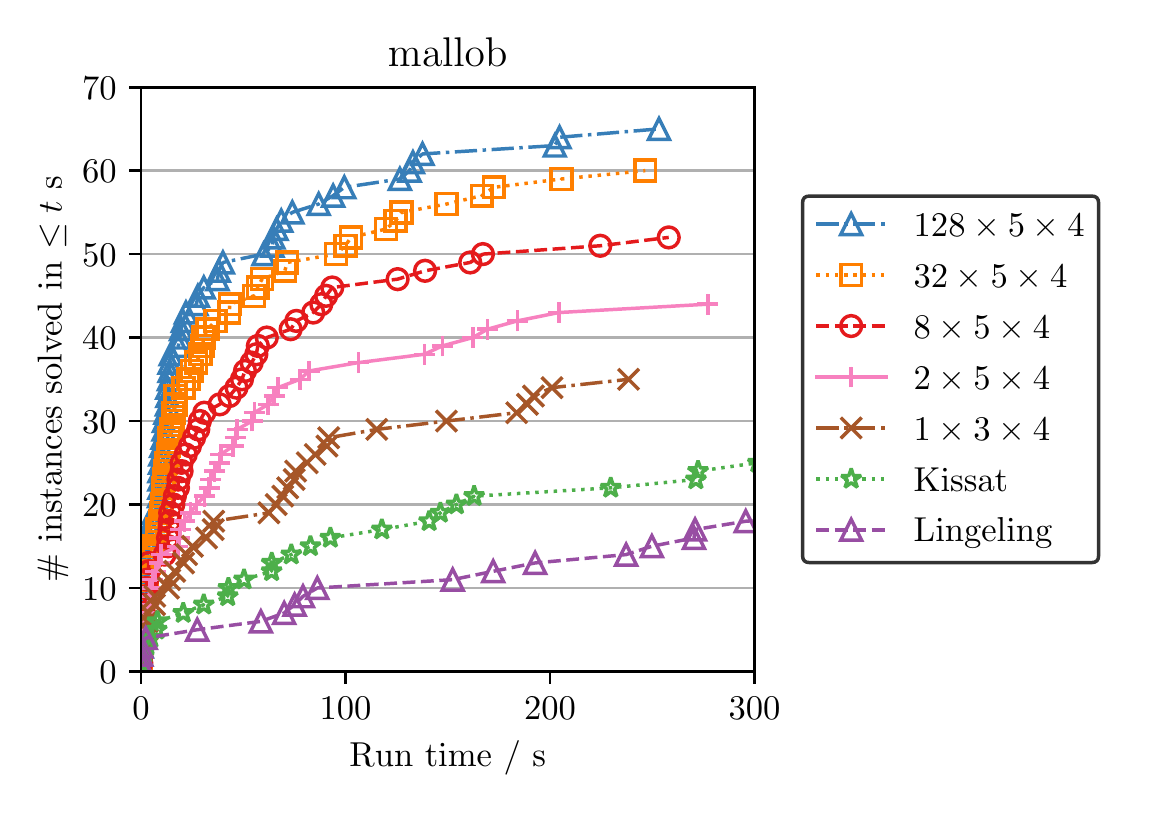}\vspace*{-0.3cm}
	\caption{Scaling behavior of Hordesat (with updated solvers) and Mallob ($\alpha=7/8$, without any clause length or LBD limits) compared to two sequential solvers.}
	\label{fig:scaling}
\end{figure}

\begin{table}[b!]
	\centering
	\setlength\tabcolsep{1mm}
	\scriptsize \begin{tabular}{l|r||r|r||r|r||r|r|r||r|r|r||}
		& & \multicolumn{4}{c||}{All instances} & \multicolumn{6}{c||}{Hard instances} \\
		& & \multicolumn{2}{c||}{Lingeling} & \multicolumn{2}{c||}{Kissat} & \multicolumn{3}{c||}{Lingeling} & \multicolumn{3}{c||}{Kissat} \\
		Config. & \# & $S_{med}$ & $S_{tot}$ & $S_{med}$ & $S_{tot}$ & \# & $S_{med}$ & $S_{tot}$ & \# & $S_{med}$ & $S_{tot}$ \\ \hline \hline
		H$1{\times}3{\times}4$ & 36 & 3.84 & 51.90 & 2.22 & 29.55 & 32 & 4.39 & 52.01 & 31 & 4.03 & 32.49 \\
		H$2{\times}5{\times}4$ & 40 & 12.00 & 95.80 & 5.06 & 64.44 & 35 & 12.27 & 96.83 & 33 & 9.11 & 69.63 \\
		H$8{\times}5{\times}4$ & 49 & 22.83 & 135.55 & 9.76 & 90.08 & 38 & 32.00 & 142.76 & 32 & 24.88 & 105.94 \\
		H$32{\times}5{\times}4$ & 56 & 42.12 & 203.66 & 15.25 & 112.14 & 34 & 97.61 & 231.77 & 19 & 114.86 & 208.68 \\
		H$128{\times}5{\times}4$ & 59 & 50.35 & 204.10 & 17.38 & 111.46 & 21 & 356.33 & 444.12 & 10 & 243.42 & 375.04 \\\hline
		M$1{\times}3{\times}4$ & 35 & 4.83 & 58.15 & 3.62 & 64.66 & 31 & 5.37 & 58.24 & 30 & 5.29 & 66.08 \\
		M$2{\times}5{\times}4$ & 44 & 12.98 & 94.44 & 10.52 & 67.71 & 39 & 14.37 & 95.28 & 37 & 11.54 & 69.25 \\
		M$8{\times}5{\times}4$ & 52 & 28.38 & 154.62 & 12.06 & 89.61 & 41 & 34.29 & 162.23 & 34 & 23.43 & 106.85 \\
		M$32{\times}5{\times}4$ & 60 & 53.75 & 220.92 & 23.41 & 148.57 & 37 & 152.19 & 245.54 & 23 & 134.07 & 262.04 \\
		M$128{\times}5{\times}4$ & 65 & 81.60 & 308.48 & 25.97 & 175.58 & 25 & 363.32 & 447.97 & 12 & 363.32 & 483.11 \\
	\end{tabular}
	\vspace{0.3cm}
	\caption{\vorschlag{Parallel speedups for Hordesat (H) and Mallob (M). In the left half, ``\#'' denotes the number of instances solved by the parallel approach and $S_{med}$ ($S_{tot}$) denotes the median (total) speedup for these instances compared to Lingeling / Kissat. In the right half, only instances are considered for which the sequential solver took at least (num. cores of parallel solver) seconds to solve. Here, ``\#'' denotes the number of considered instances for each combination.}}
	\label{tab:speedups}
\end{table}
We now present some scaling results.
Fig.~\ref{fig:scaling} provides an overview on the performance of both Hordesat and Mallob when executed on 12, 40, 160, 640, and 2560 cores.
As sequential baselines we included Lingeling (in the 2018 version used by Mallob) as well as Kissat \cite{biere2020cadical}, the winner of the SAT Competition 2020's main track.
\vorschlag{Table~\ref{tab:speedups} shows pairwise speedups.
We used a time limit of $T_s = 50~000s$ for sequential solvers and $T_p = 300s$ for parallel solvers.
As in \cite{balyo2015hordesat} we ``generously'' attribute a run time of $T_s$ to the sequential approach for each unsolved instance solved by the parallel approach.
We computed the median speedup $S_{med}$ and the total speedup $S_{tot}$ (the sum of all sequential run times divided by the sum of all parallel run times).
We also provide speedups emulating ``weak scaling'', i.e., only considering instances for which the sequential approach took at least (\# cores of parallel approach) seconds.}

While both parallel solvers show noticeable speedups whenever the amount of resources is (approximately) quadrupled, the 128-node configuration of Hordesat shows clear signs of diminishing returns and only performs slightly better than its 32-node configuration.
As such, Mallob on only 32 nodes outperforms Hordesat on 128 nodes.
Furthermore, the 128-node configuration of Mallob achieves a much more pronounced speedup over its 32-node configuration when compared to Hordesat, although we do notice some degree of diminishing returns as well.
\vorschlag{This decline in efficiency motivates the next stage of our evaluations where we employ Mallob to resolve multiple jobs in parallel.}

\subsection{Malleable Job Scheduling}

To evaluate Mallob in its entirety, we made use of 128 compute nodes as follows:
One PE is a designated ``client'' which introduces jobs to the system and receives results or timeout notifications.
We limit the maximum number $J$ of active jobs in the system to 4, 16, and 64 respectively.
Furthermore, the randomized scheduling and load balancing paradigm of Mallob requires 
that a small ratio $\varepsilon$ of PEs is reserved to remain idle.
We chose $\varepsilon = 0.05$ which we expect to be on the conservative side (i.e., lower values of $\varepsilon$ can still be viable).
\vorschlag{We partition each compute node into 4 PEs with 5 threads each because this better fits the two-socket hardware for the additional OS-level work and communication we perform.
As such,} $\lfloor (1-\varepsilon) (p-1) \rfloor = \lfloor 0.95 \cdot 511 \rfloor = 485$ PEs should actively perform SAT solving at any given time.
For $J=4$ (16, 64), this leads to around 121 (30, 7) PEs or 605 (150, 35) threads per job compared to the 640 (160, 40) threads of the closest tested standalone configuration of Mallob.
\vorschlag{We limited the number of job nodes residing on any PE at the same time to three.}

\begin{figure}[b]
	\centering
	\begin{subfigure}{0.54\textwidth}
		\includegraphics[width=\textwidth]{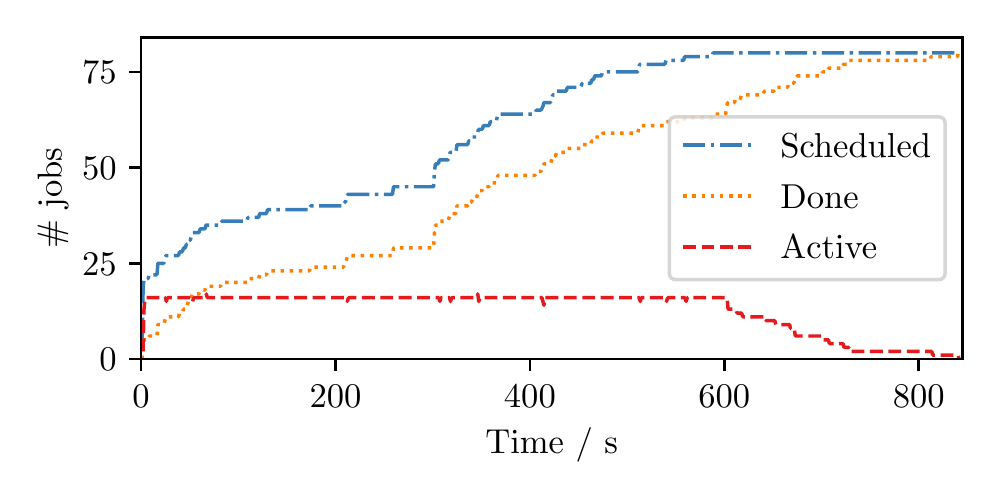}	
	\end{subfigure}
	\begin{subfigure}{0.45\textwidth}
		\setlength\tabcolsep{1mm}
		\scriptsize \begin{tabular}{l|rrrr}
			Approach & \# & (+, & -) & PAR-2 \\\hline\hline
			Mallob $J=4$ & 58 & 26 & 32 & 192.7 \\
			Mb-mono $m=32$ & \textbf{60} & \textbf{28} & 32 & \textbf{181.4} \\\hline
			Mallob $J=16$ & \textbf{54} & \textbf{24} & \textbf{30} & \textbf{232.7} \\
			Mb-mono $m=8$ & 52 & 23 & 29 & 240.1 \\\hline
			Mallob $J=64$ & \textbf{49} & \textbf{21} & \textbf{28} & \textbf{279.0} \\
			Mb-mono $m=2$ & 44 & 19 & 25 & 299.8 \\
		\end{tabular}
	\end{subfigure}
	\caption{\vorschlag{Left: Number of active jobs and cumulative number of scheduled and done jobs while processing 80 jobs with $J=16$, measured at a resolution of 1 second.
	Right: Solved instances (SAT, UNSAT) and PAR-2 scores of Mallob with $J=4,16,64$ and of similar configurations of Mallob-mono.}}
	\label{fig:sched-load-jobs}
	\label{tab:performance-scheduled-vs-mono}
\end{figure}

\vorschlag{For 96\% of all measurements we counted exactly 485 busy PEs (94.9\% system load).}
As Fig.~\ref{fig:sched-load-jobs} illustrates for $J=16$, the number of jobs scheduled so far and the number of done jobs progress uniformly until no more jobs are left:
The job scheduling times (measured from the introduction of the initial job request $r_0(j)$ to the initiation of the transfer of the job description to $p_0(j)$) ranged from 0.003~s to 0.781~s (average 0.061~s, median 0.006~s).
Our scheduling and load balancing imposes very little overhead: With $J=4$ (16, 64) we have measured an average of 3.1\% (3.0\%, 3.0\%) of active core time in the PEs' main threads which collectively perform the entire scheduling, load balancing, and communication.
We measured an average CPU usage of 99.1\% (99.2\%, 99.0\%) for the SAT solver threads.
To measure the quality of our resource allocation, we compute the \textit{over-transfer factor} $f := \frac{\#starts}{\sum_j \max\{v_j\}}$ as the total number of initialized job nodes divided by the sum of each job's maximum volume.
Ideally, $f = 1$ if all $p_x(j)$ for $0 \leq x < \max\{v_j\}$ are initialized exactly once and reused whenever the volume fluctuates.
\vorschlag{We measured $f = 1.16$ (1.08, 1.002) for $J=4$ (16, 64).}

\vorschlag{Let us now compare our system with $J=4$ (16, 64) with the performance of Mallob-mono on 640 (160, 40) cores.
Table~\ref{tab:performance-scheduled-vs-mono} shows that the run with $J=4$ performed slightly worse, the run with $J=16$ performed slightly better and the run with $J=64$ performed considerably better than its respective \textit{mono} configuration:
When the number of active jobs diminishes towards the end of the run, as can be seen on the right of Fig.~\ref{fig:sched-load-jobs}, the remaining jobs receive additional computational power which has a positive effect on the overall performance.
This effect is more pronounced the more jobs are being processed overall.}

\begin{figure}[b!]
	\centering
	\begin{subfigure}{0.5\textwidth}
		\includegraphics[width=\textwidth]{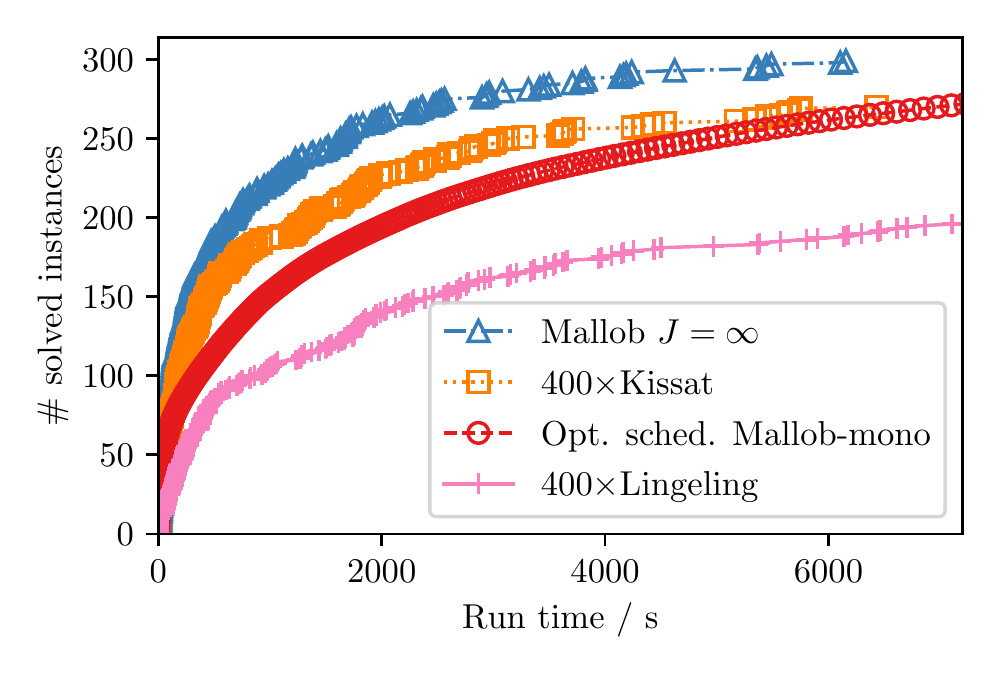}	
	\end{subfigure}
	\begin{subfigure}{0.49\textwidth}
		\setlength\tabcolsep{1mm}
		\scriptsize \begin{tabular}{l||rr|rr}
			& \multicolumn{2}{c|}{$R_{all}$} & \multicolumn{2}{c}{$R_{slv}$} \\
			Configuration & avg. & med. & avg. & med.\\\hline
			Mallob $J=\infty$ & 2422.4 & 679.8 & 808.6 & 260.6 \\
			HOS & 2914.0 & 2177.8 & 1388.5 & 913.0 \\
			$400\times$Kissat & 2998.4 & 1362.5 & 975.5 & 355.5 \\
			$400\times$Lingeling & 4436.0 & 7200 & 1559.2 & 819.9 \\
		\end{tabular}
	\end{subfigure}
	\caption{\vorschlag{Left: Cumulative solved instances by different scheduling approaches on 128 compute nodes within two hours. Right: Average and median response times, calculated for all 400 instances ($R_{all}$) and for the solved instances per approach ($R_{slv}$). Each unsolved instance leads to a response time of 7200s.}}
\end{figure}
\vorschlag{Next, we demonstrate a more extreme use case for our system which cannot be  approximated with rigid scheduling.
We use the same system setup as described above.
We aim to resolve as many instances among the entire benchmark set of the SAT Competition 2020 (400 instances) in a limited amount of time.
Therefore, we introduce all 400 jobs at once at system start and do not impose any time limit per job.
As a comparison, we use the measured performance of our standalone SAT solving engine and wrap it in a hypothetical optimal scheduler (HOS) which has perfect a-priori knowledge of each job's run time:
To obtain optimal average response times, the jobs are scheduled sorted by their run time in ascending order.
We also include trivial schedulers which process all jobs ``embarrassingly parallel'' by running 400 instances of Lingeling or Kissat. 
}

\vorschlag{While the HOS of Mallob-mono outperforms the embarrassingly parallel version of its solver backend Lingeling, embarrassingly parallel Kissat dominates the HOS over almost the entire time frame we tested (solving a single instance less overall), which underlines both the great performance of Kissat and the high resource efficiency of (state-of-the-art) sequential SAT solvers.
However, Mallob in its true scheduling mode outperforms any of the extremes as it combines parallel job processing with a flexible degree of parallel SAT solving.
After all jobs were introduced, the average number of cores per job increased steadily from 7.2 to 24 as more and more jobs finished.
Our system solves 299 instances within 4378 core hours while the HOS solves 295 instances within 8102 core hours.
To put these measures in perspective \cite{balyo2020competition}, the solver which solved the most instances in the entirety of the SAT Competition 2020, namely a preliminary version of Mallob-mono, spent 29449 core hours (Ch) for solving the same number of 299 instances (7005Ch for solved instances, 22444Ch for unsolved instances) 
on processors which we estimate to be comparable\footnote{\vorschlag{Cloud track hardware: AWS \textit{m4.4xlarge} machines with Intel Xeon E5-2676 v3}}.
The winning system of the parallel track solved 284 instances within 6548Ch (1392Ch for solved and 5156Ch for unsolved instances) on processors we estimate to perform slightly better\footnote{\vorschlag{Parallel track hardware: AWS \textit{m4.16xlarge} machines with Intel Xeon E5-2686 v4}}.

All in all, our system is able to find a flexible trade-off between the resource-efficiency of parallel job processing and the speedups obtained by parallel SAT solving based on the current system load.
For real world applications, various mechanisms of Mallob can help to steer this degree of parallelism, such as limiting the maximum number $J$ of concurrent jobs, setting individual job priorities, and limiting a job's maximum volume and its wallclock or CPU time budget.}

\section{Conclusion}
\label{sec:conclusion}

\vorschlag{In order to improve the scalability and resource-efficiency of on-demand SAT solving in cloud environments, we outlined a framework for the scalable resolution of SAT jobs.
We presented a novel SAT solving engine based on Hordesat which incorporates communication-efficient clause exchange, a reworked solver backend supporting malleability, and various practical improvements.}
We showed that our standalone SAT solver empirically outperforms an improved Hordesat and leads to better speedups. 
We observed that our job scheduling and load balancing involves very low overhead \vorschlag{and that Mallob's combination of parallel job processing and flexible parallel SAT solving is able to reduce scheduling and response times and maximize resource efficiency in a cloud environment.}

While we focused on Mallob's SAT solving capabilities in this work, for future work we intend to evaluate the general scheduling and load balancing properties of Mallob under realistic (e.g., Poisson-distributed) job arrival rates and varying job priorities.
Secondly, while Mallob's performance with just a single solver backend is a remarkable result, we expect to further improve Mallob's performance by integrating further state-of-the-art solver backends.
Thirdly, we intend to advance Mallob by adding support for incremental SAT solving and for related applications such as automated planning \cite{schreiber2021lilotane}. 

%
%
%
\bibliographystyle{splncs04}
\bibliography{bibliography.bib}

\begin{thebibliography}{10}
\providecommand{\url}[1]{\texttt{#1}}
\providecommand{\urlprefix}{URL }
\providecommand{\doi}[1]{https://doi.org/#1}

\bibitem{audemard2014dolius}
Audemard, G., Hoessen, B., Jabbour, S., Piette, C.: Dolius: A distributed
  parallel {SAT} solving framework. In: POS@ SAT. pp. 1--11. Citeseer (2014)

\bibitem{audemard2016adaptive}
Audemard, G., Lagniez, J.M., Szczepanski, N., Tabary, S.: An adaptive parallel
  {SAT} solver. In: International Conference on Principles and Practice of
  Constraint Programming. pp. 30--48. Springer (2016)

\bibitem{audemard2009predicting}
Audemard, G., Simon, L.: Predicting learnt clauses quality in modern {SAT}
  solvers. In: Twenty-first International Joint Conference on Artificial
  Intelligence. pp. 399--404 (2009)

\bibitem{balyo2020proceedings}
Balyo, T., Froleyks, N., Heule, M.J., Iser, M., J{\"a}rvisalo, M., Suda, M.:
  Proceedings of {SAT} competition 2020: Solver and benchmark descriptions
  (2020)

\bibitem{balyo2020competition}
Balyo, T., Froleyks, N., Heule, M.J., Iser, M., J{\"a}rvisalo, M., Suda, M.:
  {SAT} competition (2020), \url{https://satcompetition.github.io/2020/},
  accessed: 2021-03-19.

\bibitem{balyo2015hordesat}
Balyo, T., Sanders, P., Sinz, C.: Hordesat: A massively parallel portfolio
  {SAT} solver. In: International Conference on Theory and Applications of
  Satisfiability Testing. pp. 156--172. Springer (2015)

\bibitem{biere2014yet}
Biere, A.: Yet another local search solver and lingeling and friends entering
  the {SAT} competition 2014. Sat competition  \textbf{2014}(2), ~65 (2014)

\bibitem{biere2016splatz}
Biere, A.: Splatz, lingeling, plingeling, treengeling, yalsat entering the
  {SAT} competition 2016. Proc. of SAT Competition pp. 44--45 (2016)

\bibitem{biere2017cadical}
Biere, A.: Cadical, lingeling, plingeling, treengeling and yalsat entering the
  {SAT} competition 2018. Proceedings of SAT Competition pp. 14--15 (2017)

\bibitem{biere2020cadical}
Biere, A., Fazekas, K., Fleury, M., Heisinger, M.: {CaDiCaL}, kissat,
  paracooba, plingeling and treengeling entering the {SAT} competition 2020.
  SAT COMPETITION 2020 p.~50 (2020)

\bibitem{buning2019using}
B{\"u}ning, M.K., Balyo, T., Sinz, C.: Using {DimSpec} for bounded and
  unbounded software model checking. In: International Conference on Formal
  Engineering Methods. pp. 19--35. Springer (2019)

\bibitem{ehlers2014communication}
Ehlers, T., Nowotka, D., Sieweck, P.: Communication in massively-parallel {SAT}
  solving. In: 2014 IEEE 26th international conference on tools with artificial
  intelligence. pp. 709--716. IEEE (2014)

\bibitem{graham2006open}
Graham, R.L., Shipman, G.M., Barrett, B.W., Castain, R.H., Bosilca, G.,
  Lumsdaine, A.: Open {MPI}: A high-performance, heterogeneous {MPI}. In: 2006
  IEEE International Conference on Cluster Computing. pp.~1--9. IEEE (2006)

\bibitem{gropp1999using}
Gropp, W., Gropp, W.D., Lusk, E., Skjellum, A., Lusk, E.: Using MPI: portable
  parallel programming with the message-passing interface, vol.~1. MIT press
  (1999)

\bibitem{hamadi2010manysat}
Hamadi, Y., Jabbour, S., Sais, L.: {ManySAT}: a parallel {SAT} solver. Journal
  on Satisfiability, Boolean Modeling and Computation  \textbf{6}(4),  245--262
  (2010)

\bibitem{heisinger2020distributed}
Heisinger, M., Fleury, M., Biere, A.: Distributed cube and conquer with
  paracooba. In: International Conference on Theory and Applications of
  Satisfiability Testing. pp. 114--122. Springer (2020)

\bibitem{heule2011cube}
Heule, M.J., Kullmann, O., Wieringa, S., Biere, A.: Cube and conquer: Guiding
  {CDCL} {SAT} solvers by lookaheads. In: Haifa Verification Conference. pp.
  50--65. Springer (2011)

\bibitem{hoos2000local}
Hoos, H.H., St{\"u}tzle, T.: Local search algorithms for {SAT}: An empirical
  evaluation. Journal of Automated Reasoning  \textbf{24}(4),  421--481 (2000)

\bibitem{iser2019memory}
Iser, M., Balyo, T., Sinz, C.: Memory efficient parallel {SAT} solving with
  inprocessing. In: 2019 IEEE 31st International Conference on Tools with
  Artificial Intelligence (ICTAI). pp. 64--70. IEEE (2019)

\bibitem{iser2019problem}
Iser, M., Sinz, C.: A problem meta-data library for research in {SAT}.
  Proceedings of Pragmatics of SAT  \textbf{59},  144--152 (2019)

\bibitem{le2017painless}
Le~Frioux, L., Baarir, S., Sopena, J., Kordon, F.: Painless: a framework for
  parallel {SAT} solving. In: International Conference on Theory and
  Applications of Satisfiability Testing. pp. 233--250. Springer (2017)

\bibitem{marques2009conflict}
Marques-Silva, J., Lynce, I., Malik, S.: Conflict-driven clause learning {SAT}
  solvers. In: Handbook of satisfiability, pp. 131--153. ios Press (2009)

\bibitem{massacci2000logical}
Massacci, F., Marraro, L.: Logical cryptanalysis as a {SAT} problem. Journal of
  Automated Reasoning  \textbf{24}(1),  165--203 (2000)

\bibitem{ngoko2019solving}
Ngoko, Y., C{\'e}rin, C., Trystram, D.: Solving {SAT} in a distributed cloud: a
  portfolio approach. International Journal of Applied Mathematics and Computer
  Science  \textbf{29}(2),  261--274 (2019)

\bibitem{ngoko2017distributed}
Ngoko, Y., Trystram, D., C{\'e}rin, C.: A distributed cloud service for the
  resolution of {SAT}. In: 2017 IEEE 7th International Symposium on Cloud and
  Service Computing (SC2). pp.~1--8. IEEE (2017)

\bibitem{schreiber2021lilotane}
Schreiber, D.: Lilotane: A lifted {SAT}-based approach to hierarchical
  planning. Journal of Artificial Intelligence Research  \textbf{70},
  1117--1181 (2021)

\bibitem{schubert2010pamiraxt}
Schubert, T., Lewis, M., Becker, B.: {PaMiraXT}: Parallel {SAT} solving with
  threads and message passing. Journal on Satisfiability, Boolean Modeling and
  Computation  \textbf{6}(4),  203--222 (2010)

\end{thebibliography}


\begin{thebibliography}{8}
	\bibitem{ref_article1}
	Author, F.: Article title. Journal \textbf{2}(5), 99--110 (2016)
	
	\bibitem{ref_lncs1}
	Author, F., Author, S.: Title of a proceedings paper. In: Editor,
	F., Editor, S. (eds.) CONFERENCE 2016, LNCS, vol. 9999, pp. 1--13.
	Springer, Heidelberg (2016). \doi{10.10007/1234567890}
	
	\bibitem{ref_book1}
	Author, F., Author, S., Author, T.: Book title. 2nd edn. Publisher,
	Location (1999)
	
	\bibitem{ref_proc1}
	Author, A.-B.: Contribution title. In: 9th International Proceedings
	on Proceedings, pp. 1--2. Publisher, Location (2010)
	
	\bibitem{ref_url1}
	LNCS Homepage, \url{http://www.springer.com/lncs}. Last accessed 4
	Oct 2017
\end{thebibliography}

\clearpage
\section*{Appendix}

\textit{Note to reviewers: Depending on the requirements of the final version, we will move the Appendix into a separate document.}

\begin{table}[h!]
	\centering
	\setlength\tabcolsep{2mm}
	\scriptsize	\begin{tabular}{lrrrrl|rrrr}
		Solver & $m$ & $\alpha$ & $X$ & CL & LBD & \# & (+, & -) & PAR-2 \\ \hline
		Mallob & 128 & $7/8$ & 10 & -- & $2\rightarrow \infty$ & 64 & 31 & 33 & 153.4 \\
		Mallob & 128 & $7/8$ & 30 & -- & $2\rightarrow \infty$ & 63 & 31 & 32 & 155.8 \\
		Mallob & 128 & $7/8$ & 90 & -- & $2\rightarrow \infty$ & 66 & 32 & 34 & 144.6 \\\hline
		Mallob & 128 & $7/8$ & -- & -- & $2\rightarrow 8$ & 63 & 31 & 32 & 156.3 \\
		Mallob & 128 & $7/8$ & -- & -- & -- & 65 & 31 & 34 & 142.7 \\\hline
		Mallob & 128 & $7/8$ & -- & 5 & -- & 65 & 32 & 33 & 144.8 \\
		Mallob & 128 & $7/8$ & -- & 10 & -- & 66 & 32 & 34 & 138.6 \\
		Mallob & 128 & $7/8$ & 90 & -- & -- & 65 & 31 & 34 & 142.3 \\
		Mallob & 128 & $7/8$ & 90 & 10 & -- & 65 & 31 & 34 & 143.0 \\\hline
		Hordesat (new, 400 inst.) & 128 & -- & -- & -- & $2\rightarrow \infty$ & 276 & 150 & 126 & 220.3 \\
		Mallob (400 inst.) & 128 & $7/8$ & -- & -- & -- & 295 & 156 & 139 & {186.0} \\\hline
		Mallob & 8$^*$& $7/8$ & -- & -- & -- & 53 & 24 & 29 & 236.0 \\
	\end{tabular}
	\vspace{0.3cm}
	\caption{Solved instances (SAT/UNSAT) and PAR-2 scores (lower is better) of further experiments.
		Parameters: number of compute nodes $m$, clause buffer discount factor $\alpha$ and half life $X$, clause length limit, initial and final LBD limit.
		$^*$These configurations used 4 PEs à 5 threads instead of 5 PEs à 4 threads.}
	\label{tab:results}
\end{table}

\begin{figure}
	\centering
	\includegraphics[width=0.6\textwidth]{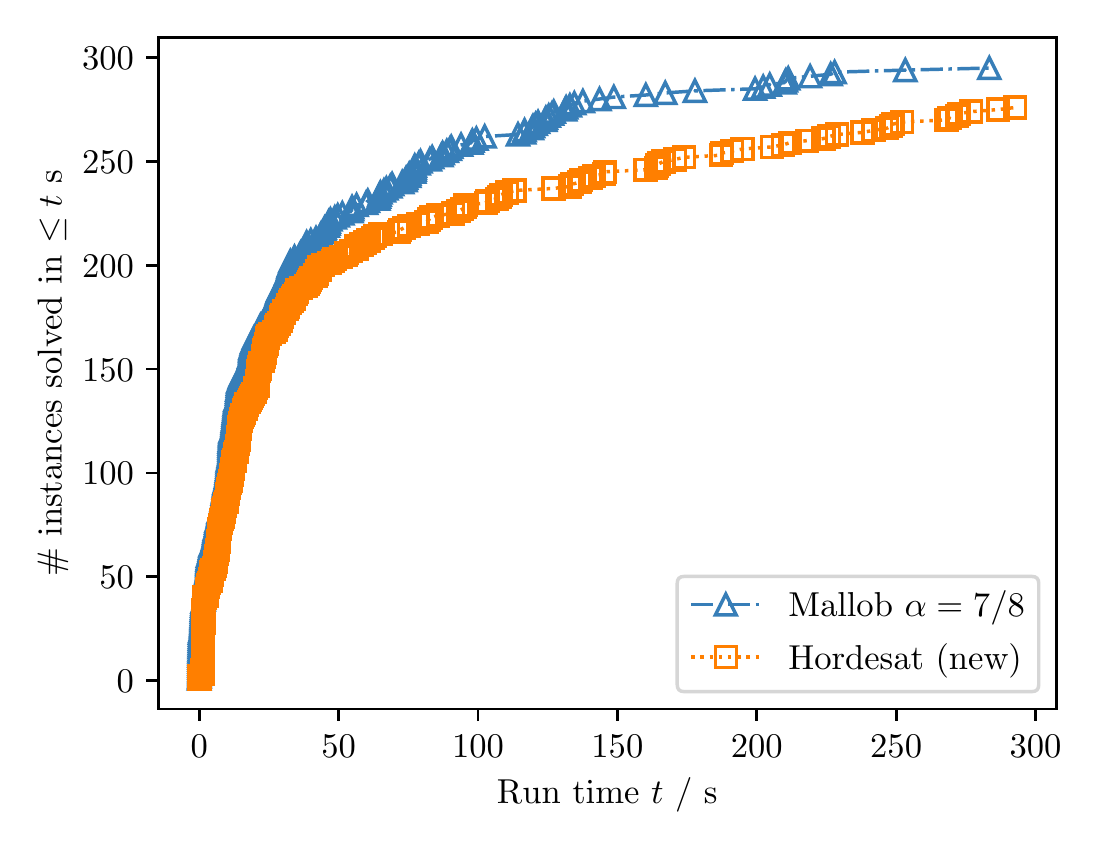}
	\caption{Performance of Mallob ($\alpha=7/8$, no LBD limits) and updated Hordesat on the entire benchmark set of the SAT Competition 2020}
\end{figure}

\begin{figure}
	\centering
	\includegraphics[width=0.49\textwidth]{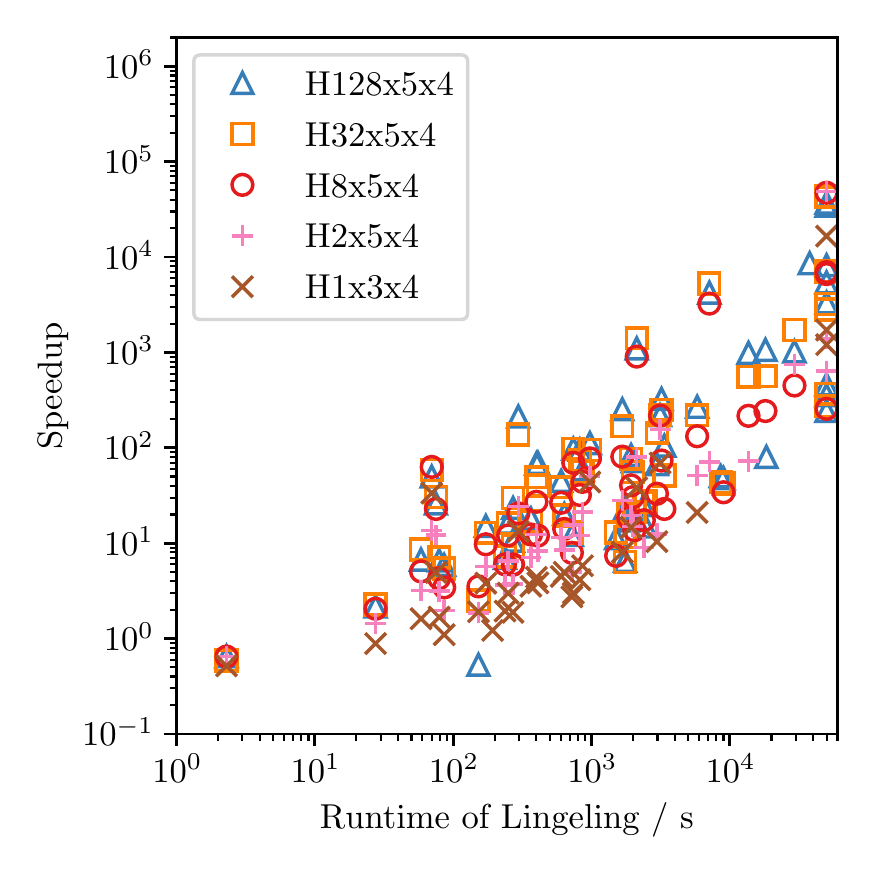}
	\includegraphics[width=0.49\textwidth]{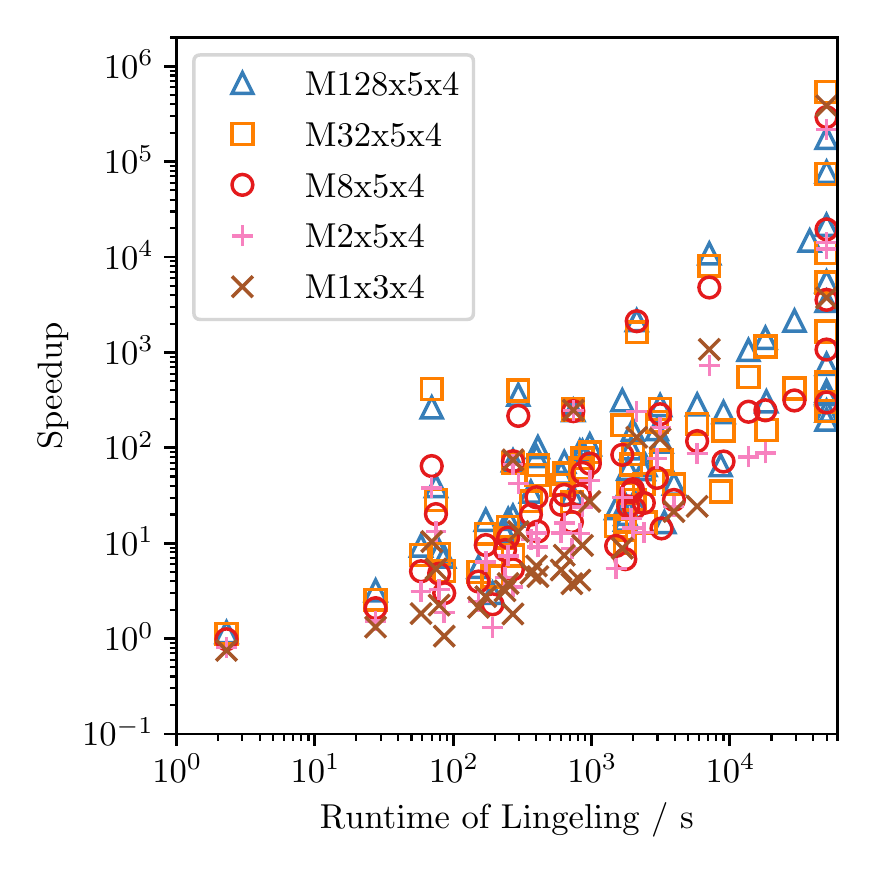}\\
	\includegraphics[width=0.49\textwidth]{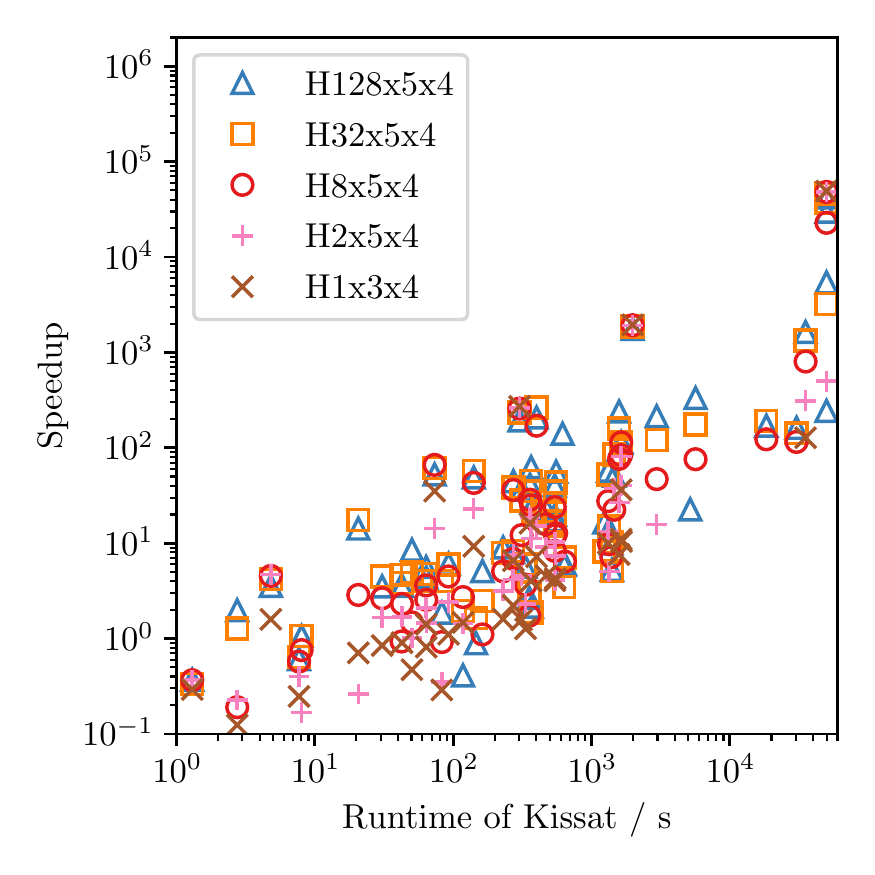}
	\includegraphics[width=0.49\textwidth]{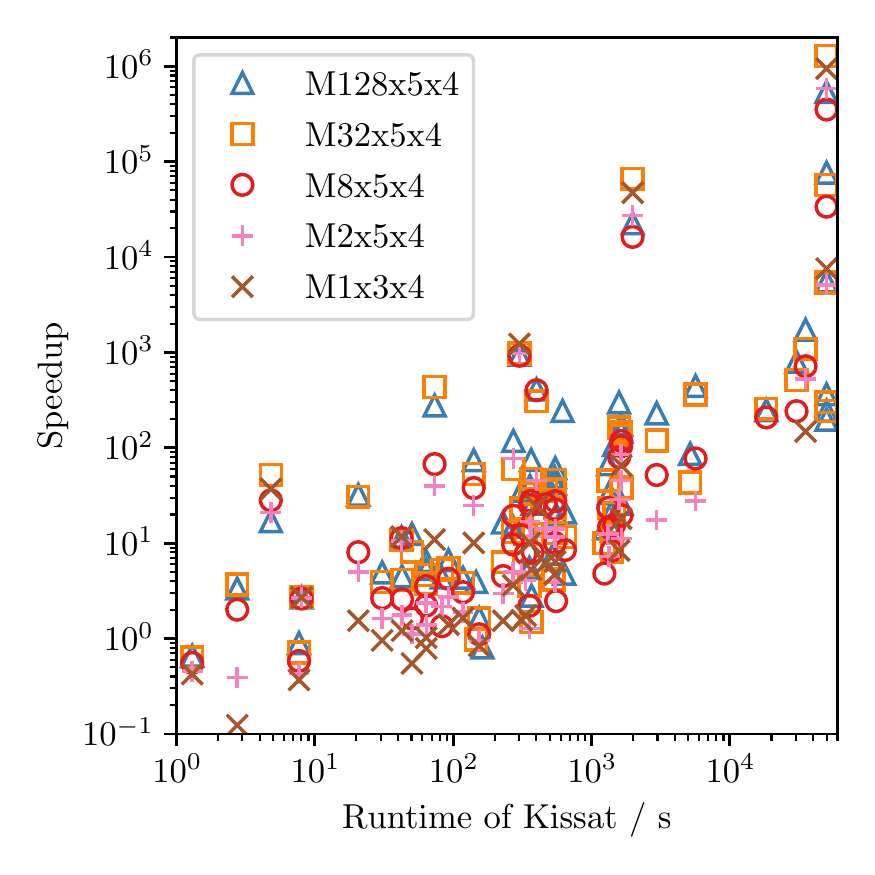}
	\caption{Parallel speedups of Mallob and Hordesat over Lingeling and Kissat, only showing sequential run times of at least one second. Points at the right border denote timeouts of the sequential solver. We omitted a single speedup of $<0.1$ (of ``H1x3x4'' over Kissat). Note the logarithmic scales.}
\end{figure}

\begin{figure}
	\centering
	\includegraphics[width=0.8\textwidth]{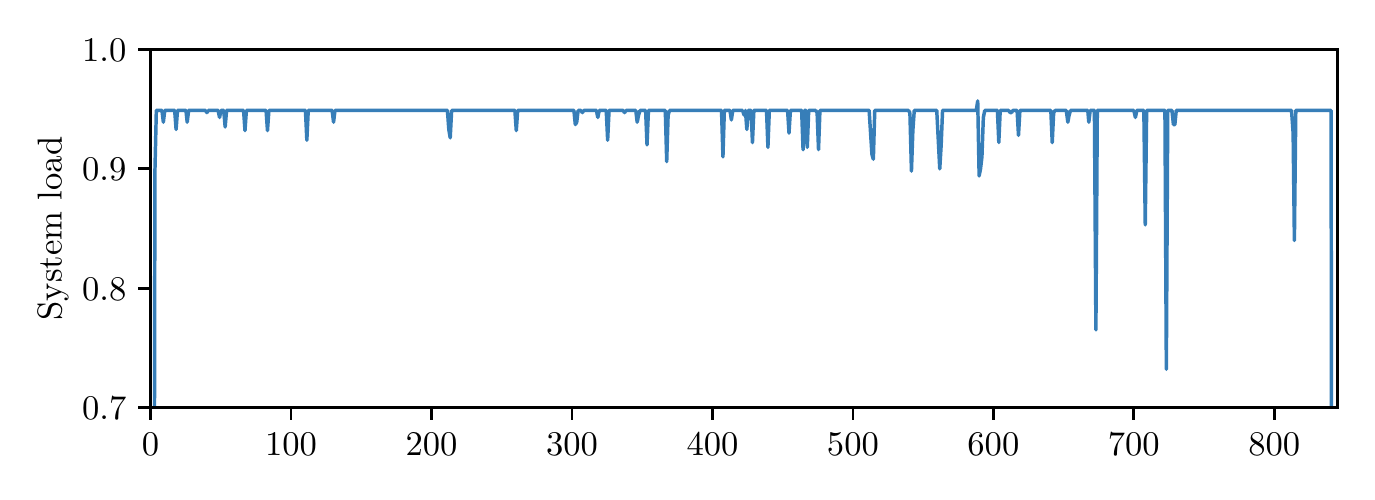}
	\caption{Ratio of busy PEs of Mallob with $J=16$, measured every second.}
\end{figure}
\begin{figure}
	\centering
	\includegraphics[width=0.6\textwidth]{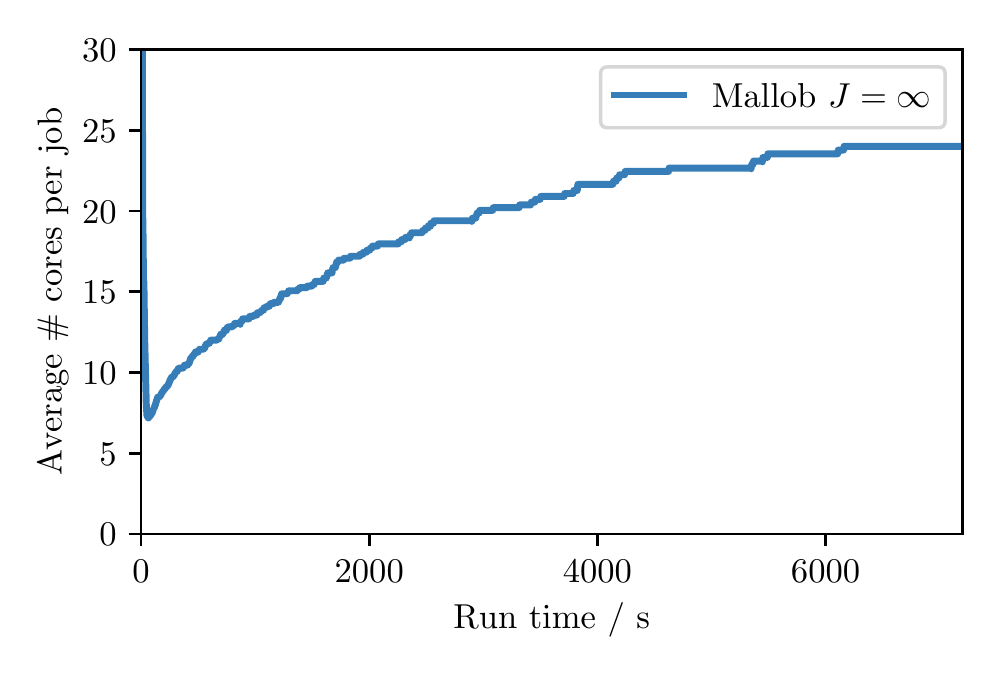}
	\caption{Active cores per job of Mallob processing 400 jobs at once}
\end{figure}

\begin{figure}
	\centering
	\includegraphics[width=0.8\textwidth]{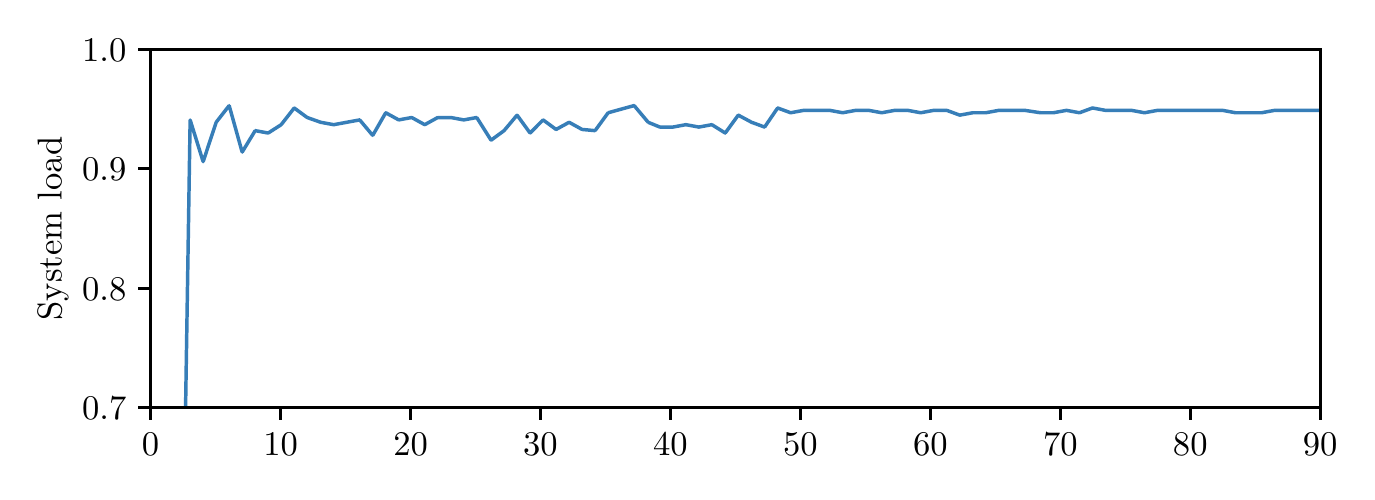}\\
	\includegraphics[width=0.8\textwidth]{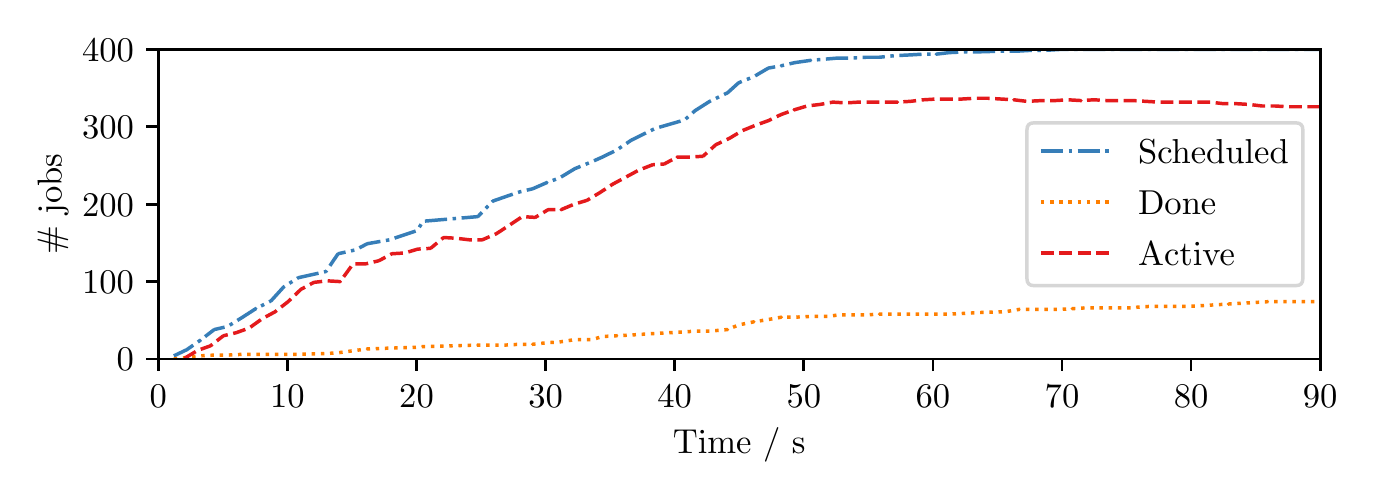}
	\caption{System load and job distribution in the first 90 seconds execution time of Mallob processing 400 jobs at once.}
\end{figure}

\end{document}